\documentclass[a4paper,
               keeplastbox,   % flushend option: not to un-indent last line in References
               nospread,     % flushend option: do not fill with whitespace to balance columns
               ]{jacow}
\usepackage{pdfpages,multirow,ragged2e} %
\usepackage[switch]{lineno}
\makeatletter%
	\ifboolexpr{bool{xetex}}
	 {\renewcommand{\Gin@extensions}{.pdf,%
	                    .png,.jpg,.bmp,.pict,.tif,.psd,.mac,.sga,.tga,.gif,%
	                    .eps,.ps,%
	                    }}{}
\makeatother

\ifboolexpr{bool{xetex} or bool{luatex}} % test for XeTeX/LuaTeX
 {}                                      % input encoding is utf8 by default
 {\usepackage[utf8]{inputenc}}           % switch to utf8

\usepackage[USenglish]{babel}
\usepackage{comment}
\usepackage{float} %required for the placement specifier H

\ifboolexpr{bool{jacowbiblatex}}%
 {%
  \addbibresource{jacow-test.bib}
  \addbibresource{biblatex-examples.bib}
 }{}
\begin{document}

\title{Controlling the CERN Experimental Area Beams}

\author{B. Rae\thanks{Bastien.Rae@cern.ch}, M. Hrabia, V. Baggiolini, D. Banerjee, J. Bernhard, M. Brugger, N. Charitonidis,\\L. Gatignon, A. Gerbershagen, R. Gorbonosov, M. Peryt, M. Gabriel, G. Romagnoli, C. Roderick\\CERN, 1211 Geneve 23, Switzerland}
	
\maketitle

\begin{abstract}
   The CERN fixed target experimental areas are composed of more than 8\,km of beam lines with around 800 devices used to define and monitor the beam parameters. Each year more than 140 groups of users come to perform experiments in these areas, with a need to control and access the data from these devices. The software to allow this therefore has to be simple and robust, and be able to control and read out all types of beam devices. This contribution describes the functionality of the beam line control system, CESAR, and its evolution. This includes all the features that can be used by the beam line physicists, operators, and device experts that work in the experimental areas. It also underlines the flexibility that the software provides to the experimental users for control of their beam line, allowing them to manage this in a very easy and independent way. This contribution also covers the on-going work of providing MAD-X support to CESAR to achieve an easier way of integrating beam optics. An overview of the on-going software migration of the Experimental Areas is also given.
\end{abstract}

\section{Introduction}
The CERN experimental areas are a complex system of beam lines and beam intercepting devices that are able to provide a large variety of different particle beams to different experiments and detector assemblies. They serve both fixed target experiments and test beams~\cite{Banerjee:2774716}. The most important aspect of these unique experimental facilities is the possibility for experimental users to control and to monitor beam parameters from dedicated terminals installed in their respective control rooms. Such parameters include the access to the experimental zones, the beam intensity via collimator settings, the magnet currents, which are defining the beam trajectory and focal properties, the particle species via the use of targets, converters and absorbers, and the instrumentation for monitoring. The beam control system is called CESAR~\cite{Baggiolini:2004kp}, which is an acronym for CERN Experimental areas Software Renovation. Through the past 10 years, CESAR has been continuously developed with new features and devices types being added. With the new secondary beams software migration project, the CESAR scope will be extended to accept optics calculations through MAD-X connectivity, and ideally also with automatic layout updates through the CERN Layout database.

\begin{figure}[htp]
    \centering
    \includegraphics[width=0.8\columnwidth, trim={12
    0 18 0}, clip]{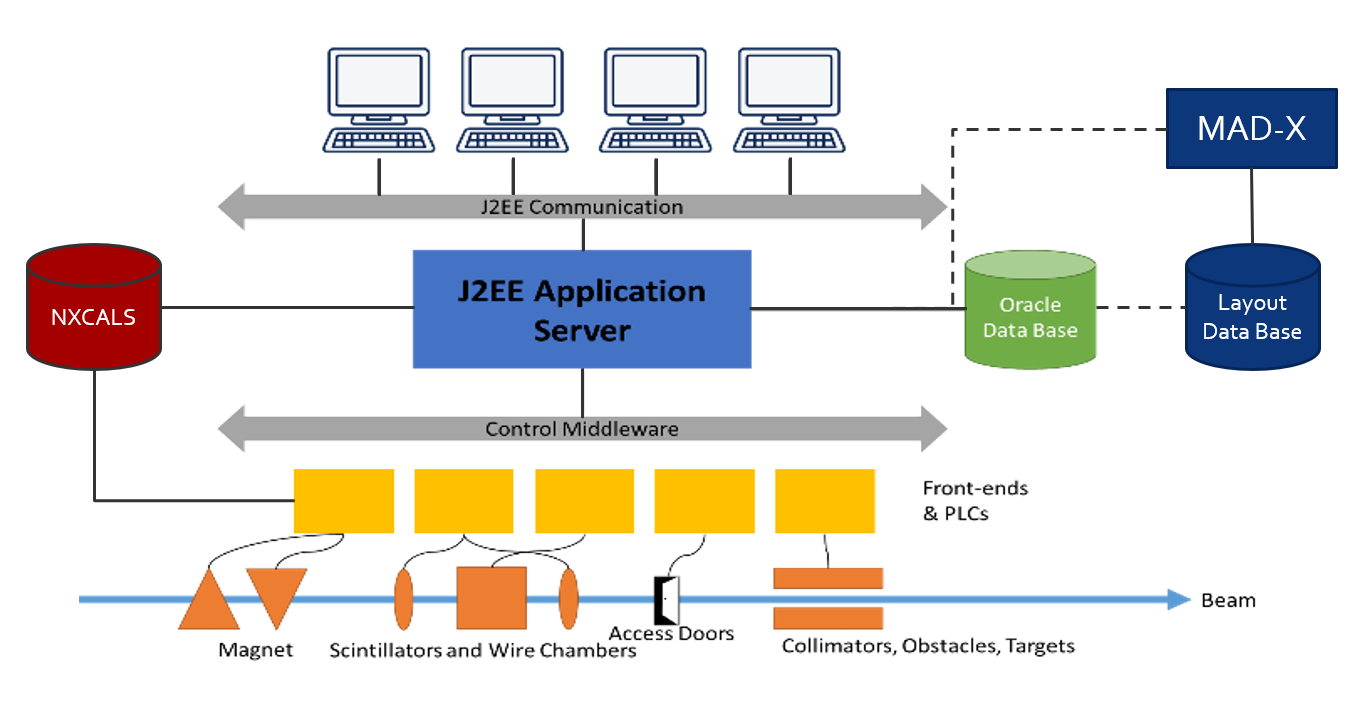}    
    \caption{CESAR Architecture and Foreseen Connectivity.}
    \label{fig:HIW_CESAR}
\end{figure}

The particularity of CESAR with respect to other control systems of the CERN accelerators is that it is designed to be operated by non-experts, as well. Many of the experimental users are not accelerator physicists and do not know all details of the beam line and its equipment. Therefore the system is made easy and intuitive, yet safe, in order to avoid any unintentional damage to the beam lines and experimental equipment. CESAR is based on Java and constructed around an ORACLE database. It acquires and sets so-called equipment knobs, mainly by subscribing to the Front-End Software Architecture FESA~\cite{FESA} device. In addition, it receives information from other services such as from the access system database (Access-DB) , via DIP (Data Interchange Protocol), and the data logging system NXCALS~\cite{Wozniak:2019tby}. All devices are identified in the CESAR database together with their parameters, such as FESA name, element type, beam line, and others. This allows flexible modifications as often needed in secondary beam lines. The architecture of CESAR is shown in Fig.~\ref{fig:HIW_CESAR}.

\section{User Types}
For both safety and simplicity reasons, there are three user types defined in the database: (1) \textit{Super Users} are allowed to see and change any parameters of all devices in all beam lines. This group is composed of the responsible beam physicists, accelerator operators, and selected equipment specialists. (2) \textit{Main Users} are associated with specific consoles in an experimental control room and are allowed to change most non-safety relevant settings in their beam line up to their experiment. They are set by the super users according to the experiment schedule, which is provided by the SPS/PS Physics Coordinator. (3) \textit{Standard Users} are treated similarly as main users, however they see only their assigned experimental area, for instance to initiate an access procedure. Standard users are able to monitor their beam parameters, but are not allowed to control any devices other than the ones in their assigned user zone.
%\begin{figure}[H]
%    \centering
%    \includegraphics[width=\columnwidth]{CESAR_users_Fig1.PNG}    
%    \caption{Main user selection panel}
%    \label{fig:User_CESAR}
%\end{figure}

\section{Interface}
\begin{figure*}[!htb]
    \centering
    \includegraphics*[width=0.85\textwidth]{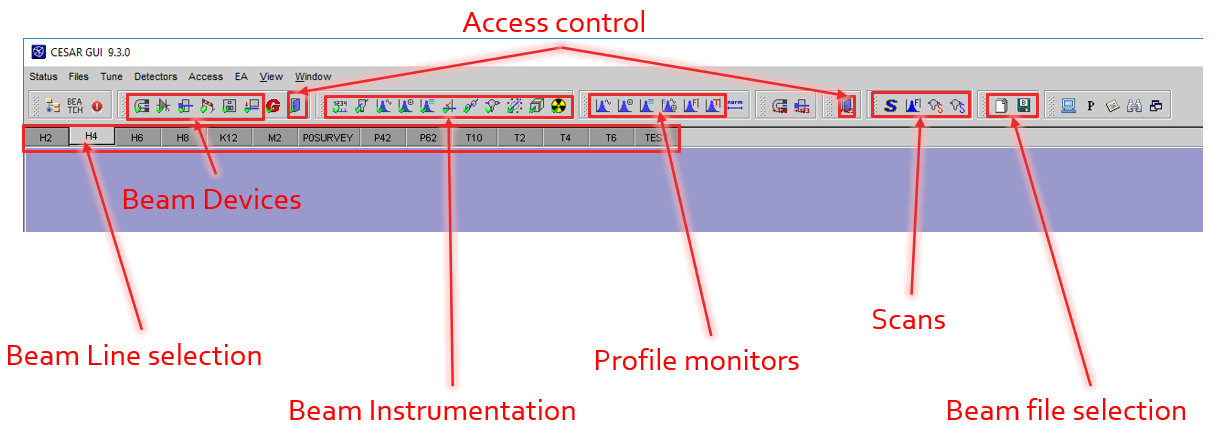}    
    \caption{CESAR interface for Super Users.}
    \label{fig:interface_CESAR}
\end{figure*}
The CESAR interface is composed of three main panels, as depicted in Fig.~\ref{fig:interface_CESAR}: the top menu, the devices panel and the beam line selection tab. The latter is used to change the selected beam line and control the devices associated with it. This functionality is reserved for super users. The devices panel displays the main features and functionality, which is needed during setting-up and tuning of beams, as well as during the operation during a physics run. On the top menu, specific panels can be opened within to the devices panel, including specific modules and panels that are related to particular beam line equipment, beam line protection, user configurations and other settings. In addition, automatic scan programs used for precise beam steering can be opened that allow efficient tuning of selected elements while visualising direct feedback by the beam instrumentation.

\section{CESAR Device Control}

\subsection{Collimators}
\begin{figure}[htb]
    \centering
    \includegraphics[width=\columnwidth]{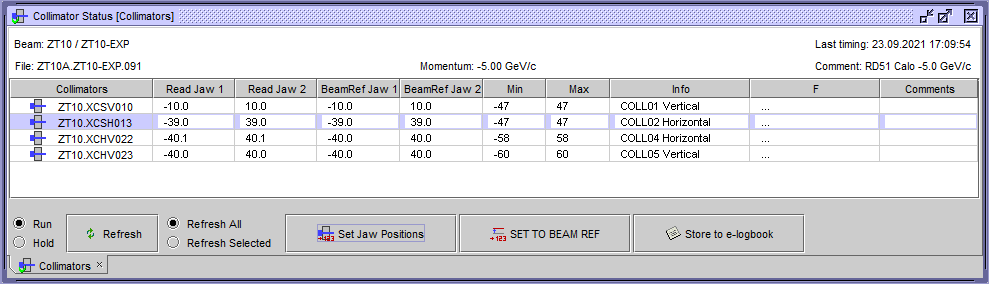} 
    \caption{Collimators Status.}
    \label{fig:coll_CESAR}
\end{figure}

For collimator settings, each of the motors moving individual jaws is controlled. Collimators with four jaws are considered as two different entities, one vertical and one horizontal, for a better overview. They are used for changing intensity, shape, and energy spread of a beam. Similar to the magnet settings, one can set reference values for each of them, as can be seen in Fig.~\ref{fig:coll_CESAR}.

\subsection{Magnets}
\begin{figure}[htbp]
    \centering
    \includegraphics[width=\columnwidth]{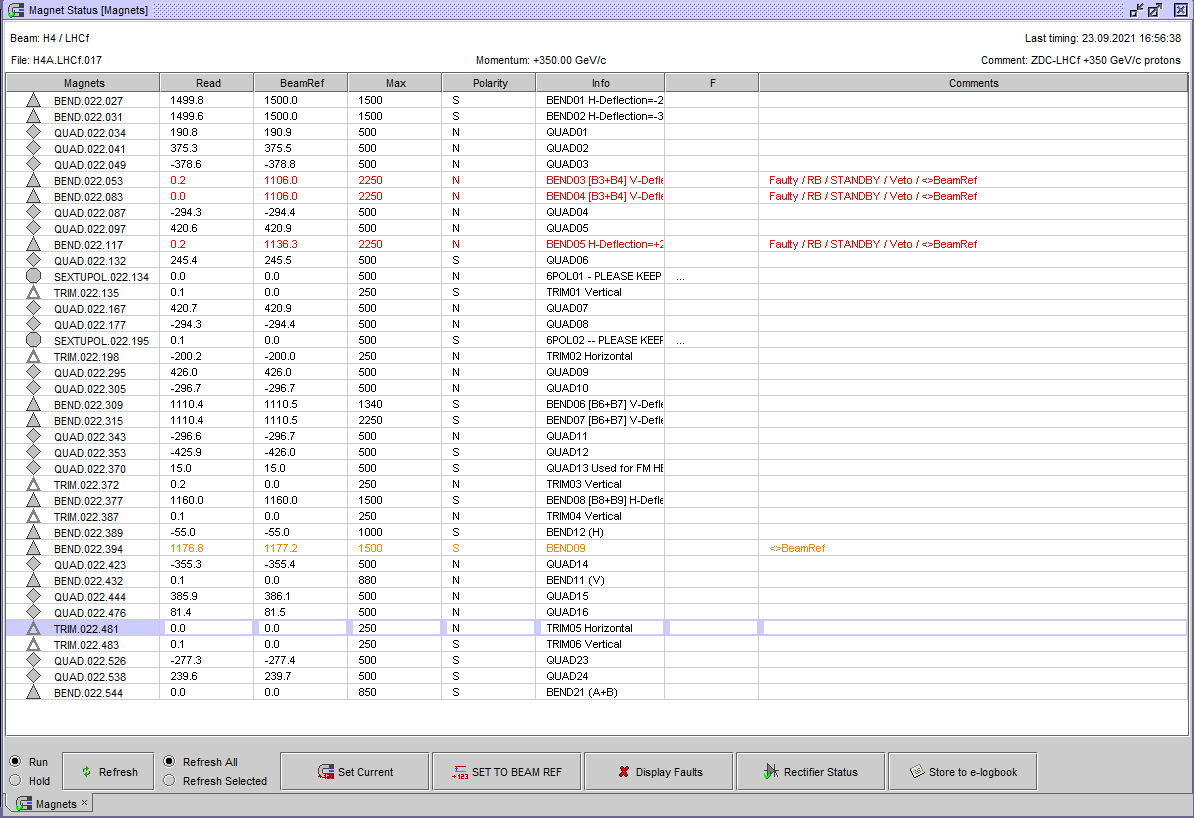}    
    \caption{Magnets Status. Note the display of faults in red colour.}
    \label{fig:magnet_CESAR}
\end{figure}

In the magnets status panel (see Fig.~\ref{fig:magnet_CESAR}), all magnets of the selected beam line are displayed together with their main parameters. There is the possibility to set and read the applied current values for each of the magnets and reference values can be defined in addition. This reference allows to go back to previous configurations, e.g. when steering the beam. CESAR also displays magnet faults together with the specific fault type, e.g. overheating. Another functionality is the so-called rectifier status, from which the power supplies can be switched on/off or moved to standby, for instance if a magnet is not in use for the currently loaded optics. It allows also resetting the power supply for certain fault types.

\subsection{Obstacles}
\begin{figure}[htb]
    \centering
    \includegraphics[width=0.6\columnwidth]{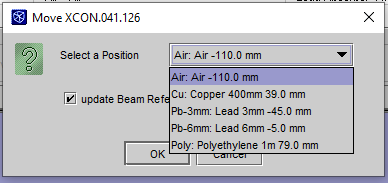} 
    \caption{Obstacles Command.}
    \label{fig:obst_CESAR}
\end{figure}

In order to create and absorb different particles types as well as for creating tertiary beams, different materials (``obstacles'') can be placed in the beam. The Obstacles Command (see Fig.~\ref{fig:obst_CESAR}) allows the users to control the position of each device and to add or remove different kinds of material automatically as these devices are motorised. The positions are all entered in the CESAR DB, so one can directly select the desired obstacle to be placed and keep a reference, as well.

\section{Beam Instrumentation}

\subsection{Scintillators and Scalers}
\begin{figure}[htp]
    \centering
    \includegraphics[width=\columnwidth]{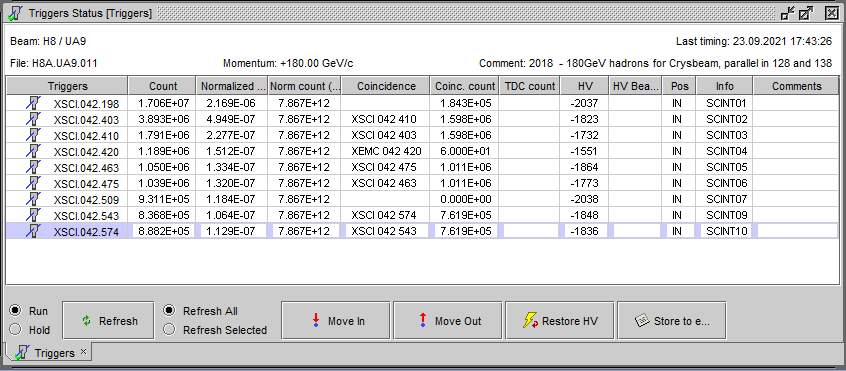} 
    \caption{Scintillator Status.}
    \label{fig:xscs_CESAR}
\end{figure}

The trigger status displays counts from each scintillator along the selected beam line, as depicted in Fig.~\ref{fig:xscs_CESAR}. In addition, it calculates ‘normalised counts’, which are normalised to the beam intensity on the upstream primary target in order to avoid fluctuations coming from the primary beam. As they are motorised, scintillators can be moved out of beam on demand, e.g. to reduce absorption for low-momentum electrons. Furthermore, in each control room, users can connect their discriminated NIM detector signals to scaler units, which are then displayed on CESAR and allow beam operators to scan and set the beam position for a maximum number of counts.

\subsection{Profile Monitors}
\begin{figure}[htp]
    \centering
    \includegraphics[width=\columnwidth]{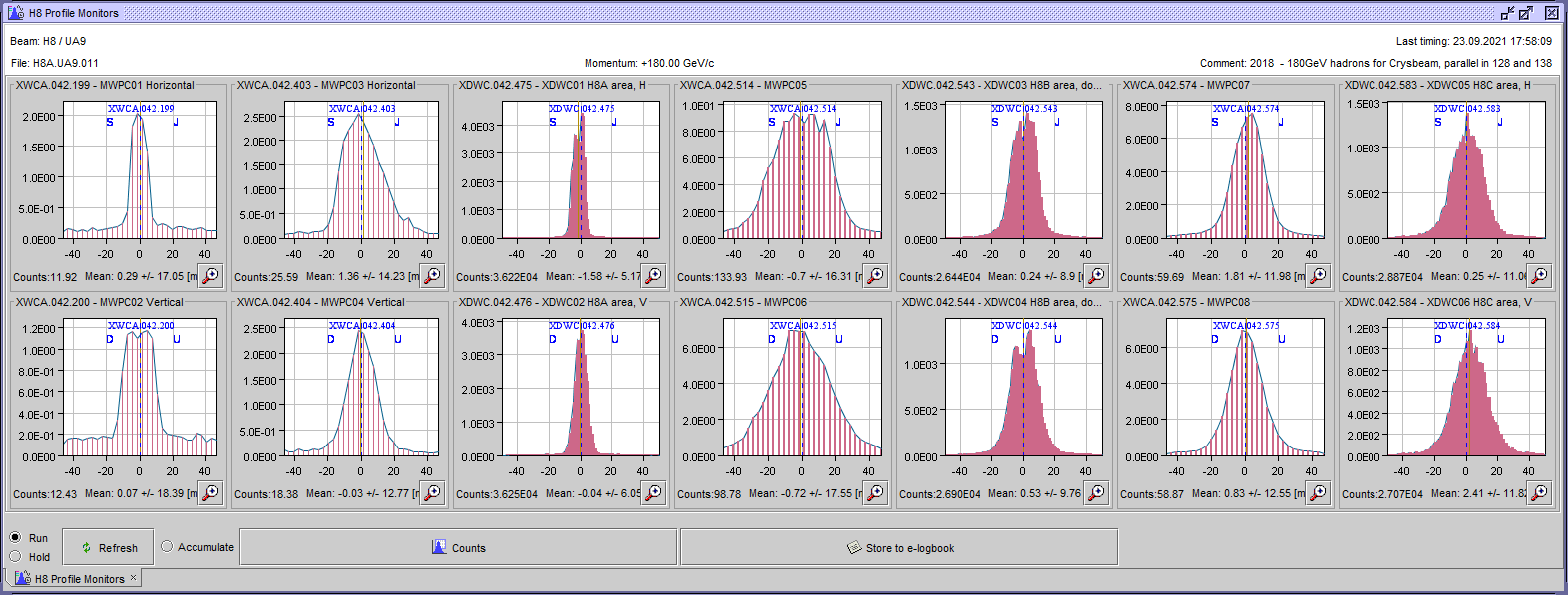} 
    \caption{Profile Monitors.}
    \label{fig:profile_monitors_CESAR}
\end{figure}

As shown in Fig.~\ref{fig:profile_monitors_CESAR}, CESAR displays beam profiles along the beam lines independent of the type of monitors that are used. Typical monitor types are analogue MWPCs, delay wire chambers and scintillating fibre monitors (XBPF). CESAR provides count rates from each monitor as well as calculated mean values of the profile distribution. As for the scintillators, some of the monitors can be moved out of the beam. Voltage settings can be adjusted by the operators for an optimal dynamic range.
\subsection{Other Instrumentation}
In addition to the above, CESAR offers displaying a variety of other beam instrumentation types specific to the selected beam line. Such detectors include the FISC scintilating wire scanners, for which CESAR reads counts versus the selected fibre position, and Cherenkov detectors for beam particle identification, for which users can set the gas pressure and even make pressure scans besides the readout of the counting rate.

\section{Improved Operational Features}
\subsection{Scans}
\begin{figure}[htp]
    \centering
    \includegraphics[width=\columnwidth]{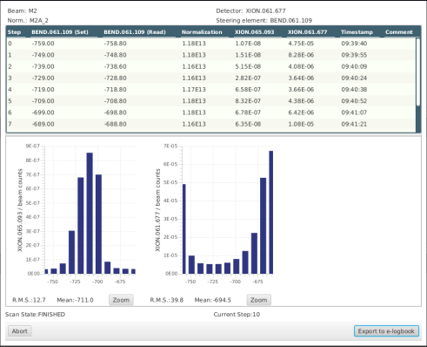} 
    \caption{Scan Result.}
    \label{fig:scan_CESAR}
\end{figure}

CESAR offers the possibility to perform scans on any beam device and instrumentation. One can select the control element (e.g. magnet or collimator) and the instrumentation to perform a scan between certain values in selected steps. The scan will go through all preset values and plot the detector reading as a function of scanned parameter, e.g. a magnet current as depicted in Fig.\ref{fig:scan_CESAR}. This needed allows to maximise transmission through a beam line or to find the position of a user detector without the need of survey in the zone. FISC scans can be performed in different modes, i.e. one position per extraction or in a fast mode for a complete scan during one extraction. There are different expert modes in addition, for instance to scan the beam divergence between two FISC monitors.

\subsection{Beam Files}
Users of experimental areas may want different beam parameters and desire different particle types, energies, and intensities. The beam files of CESAR contain all beam reference values for the selected beam lines that have been set and saved after commissioning of the specific optics and beam. This allows users to switch easily between configurations as needed by simply loading a beam file without the need of constant help of the beam operators and physicists. In addition, each file can be extrapolated to different energies, taking into account energy changes in the line as needed for tertiary beams.
%\begin{figure}[htp]
%    \centering
%    \includegraphics[width=\columnwidth]{BeamFiles.PNG} 
%    \caption{Beam Files Panel}
%    \label{fig:beam_files_CESAR}
%\end{figure}

\subsection{Access System}
Users of secondary beams typically often need access to their respective experimental areas, in particular for test beam users. It is therefore important to allow quick access to their setups without having to ask operators constantly. Hence, an access system control has been implemented in CESAR. All users can see the access status of their experimental area. The main user can both ask for access in their respective zone and turn on the beam for all users in the beam line. For safety reasons, CESAR receives information from the access system and commands the safety devices only if allowed and not prohibited by the beam interlocks. In order to make sure that no erroneous command can be sent to a device protected by the access system hardware loop, the access system matrix including all safety elements is duplicated in CESAR and verified before sending the signal. In this way, it is ensured that all conditions are fulfilled before opening a zone or turning on the beam.

%\begin{figure}[h]
%    \centering
%    \includegraphics[width=\columnwidth]{Access.PNG} 
%    \caption{The Access Status and Access Command Windows}
%    \label{fig:beam_files_CESAR}
%\end{figure}

\subsection{Radiation Monitors}
 Reading of radiation monitors has been implemented in CESAR for each beam line, in order to check the ambient radiation levels as well as to display information about the alarm thresholds. If the warning threshold is passed, the respective line will turn orange in the display window. The colour changes to red if the alarm threshold is passed. This allows a quick follow-up in case radiation alarms occur. Note that the radiation safety system is completely independent and CESAR only displays information for convenience. 

%\begin{figure}[h]
%    \centering
%    \includegraphics[width=\columnwidth]{radiation_monitors.PNG} 
%    \caption{Radiation Monitors Status}
%    \label{fig:radiation_monitors_CESAR}
%\end{figure}

\subsection{Beam Layout}
The beam layout panel displays all devices that are registered in the CESAR DB for the selected beam line. In the experimental areas, the beam lines change regularly depending on user requests. If some equipment is removed from the line, the super users can hide devices that are not needed. Those devices then are not shown in the regular status panel anymore in order to keep the overview concise. In addition, super users can ``protect'' any device, which means that any other user is inhibited from operating it anymore. Finally, Super users can add comments to any device that will be displayed in their corresponding panel, e.g. for better explanation of the device function or to highlight important features of the device setting.

%\begin{figure}[h]
%    \centering
%    \includegraphics[width=\columnwidth]{Beam_Layout.png} 
%    \caption{Beam Layout Window}
%    \label{fig:Beam_Layout_CESAR}
%\end{figure}

\section{Software Migration Project}

The Software Migration project has been initiated with the goal of modernisation of the offline software used for the description and design of the secondary beam lines at CERN. The situation at the beginning of the project in 2017, the reasons for undertaking the migration, and the baseline software  after the completion of the migration by the time of the restart of the Experimental Areas beam lines after the Long Shutdown 2 have been described in \cite{Gerbershagen:2019ueg}. The present contribution summarises the project status and puts a particular emphasis on the coupling of the software used for the beam optics calculations (MADX and AppLE.py) with CESAR.

\subsection{Project Status}
A migration of the complete software chain used for the design of the secondary beam lines in both CERN North and East Areas has been performed. The new baseline consists of MADX~\cite{MADX} for beam optics and survey calculations, the in-house developed software AppLE.py for graphical output and matching, as well as FLUKA~\cite{FLUKA1,FLUKA2,FLUKA3}, BDSIM~\cite{BDSIM} and Geant4~\cite{Geant4}, respectivel Geant4-based derivatives, for beam-matter interactions. The solution has been validated with the help of benchmark studies and a test of the complete software chain. It is planned to use the software in a highly integrated way, utilising the modern online database tools available at CERN, such as Layout Database and GitLab. 

The new software has become the baseline for Run 3 (2021 - 2025), which is expected to allow the final validation of its practicality and to reveal some aspects requiring improvement. While the major migration work has been completed according to the initial project plan, the work on the adaptation of the software chain to the evolving software infrastructure at CERN as well as the integration of the beam lines into the Layout Database, the benchmarking studies and work on further automatisation are foreseen to continue during Run 3 and beyond.

\subsection{Layout Database}
The Layout Database \cite{LayoutDB} is a CERN-wide database, designed to contain integration and installation layout data, a naming portal, photographs and drawings of the beam lines, tunnels, areas, as well as tables with all parameters relevant for the beam line description for the CERN accelerator complex. The secondary beamlines are currently being included in the database in the framework of the CERN-wide Configuration Management. It is planned to import beam line parameters such as magnet names, magnetic lengths, apertures, mapping of magnetic field strength to currents and vice versa as well as others from the CESAR database into the Layout Database and vice versa. For the latter, there exists a function of automatic generation of MADX input files from the Layout Database, which has been adapted to match the format and naming convention as required. This application takes various parameters from the database and constructs the MADX input in form of a sequence file the given beam line. This tool has been successfully tested with the K12 beam line and the sequence file has been validated successfully with the help of the previously used software.
%, the latter will also potentially have an interface with the CESAR control system software, allowing a smooth transition of the simulated settings into a beam file. 
Many of the use cases for the new software chain will be tested now thanks to the restart of beams after LS2. A large share of the North Area beam lines has still to be implemented into the Layout Database, which is planned to be completed by the end of 2021.

%\subsection{Tests and Validation}
%The aforementioned tools have been tested individually. In order to assure a smooth interfaces between the tools, a major part of the new software chain has been tested on a hardware modification on the North Area beam line K12. This modification consisted of moving several bending magnets and installing new detectors for the NA62 experiment. The following steps have been undertaken for the test: (1) BEATCH~\cite{BEATCH} files for the configurations of 2018 and 2021 have been forwarded to the EA configuration managers. (2) The configuration managers implemented those two layout versions into the Layout Database. (3) The tool for automatic production of MADX input files from the Layout Database configuration has been successfully used. (4) The file generated with the tool has been processed in MADX, producing a Survey file for the geometers. (5) The Survey file has been forwarded to the Metrology Group, which examined it and found it to be consistent with the existing Metrology Database entries with regards to naming convention and positions of the beam line elements.

\subsection{Envisaged Future Steps}

Continuing the integration of the North Area beam lines into the Layout Data Base and the MADX sequence file generation for each of them. In the medium-term, it is envisaged to create an interface between AppLE.py and CESAR.The first steps for such an integration have been taken already, allowing beam files from the CESAR database to be read by AppLE.py. That way the beam optics for any specific beam file can be calculated and visualised, allowing for instance to predict losses at collimator apertures. It is also planned to feed back newly generated and modified optics settings from AppLE.py to CESAR beam files

%The interface allows to receive an overview of the currently applied settings of the beam and provides useful insights for beam tuning. . Security aspects will have to be well considered, e.g., permitting only logged-in users with access rights to CESAR to modify beam settings. For this purpose, an Application Programming Interface (API) for CESAR would be required, as well as the integration CERN Role Based Access Control (RBAC) framework~\cite{RBAC}, providing a secure log-in into AppLE.py.This implementations are also discussed in the scope of the North Area Consolidation (NA-CONS) project.

\section{CESAR Future}
The most important aspect from the configuration management point-of-view will be the connection of CESAR to the newly commissioned beam software. The project is on a good track and several new features for CESAR have been already developed, such as the Apple.py-to-CESAR conversion and the automatic layout update with the Layout Data Base. We are thankful for the plenitude of ideas reaching us from the user community and from the recently established North Area Consolidation Project, which are evaluated at the moment. A frequently wished for item is establishing an Application Programming Interface (API) for CESAR, permitting Super Users to access the CESAR functionality from within scripts. This would allow to automatise even complicated steps for beam tuning with direct feedback from the beam instrumentation. In addition, connecting CESAR to the NXCALS logging service will allow users to retrieve recorded values of any device in convenient way. Thinking further ahead, integrating fault reporting into CESAR, e.g. with the already existing Automatic Fault System AFT~\cite{Roderick:2018pql}, will improve reliability analyses and save time of the operators. 

In addition, the new CERN GUI Strategy working group currently reviews the existing GUI systems with the aim of streamlining and easier maintainability. This is a good opportunity to improve the graphical interface and to explore possible synergies with the other control systems of CERN, for instance by adding some useful features that have been developed for accelerator controls.

\section{CONCLUSION}
CESAR is a versatile and flexible control software that is used in the experimental areas of CERN allowing users the operation of all beam devices in the secondary beam lines. It features personalised settings, such as beam configuration files, which enables quick changes of beam parameters up to a complete change of particle species and beam momentum. CESAR is being improved continuously with new features becoming available and following the evolution of users requirements. Recently, in the framework of the secondary beam software migration project, a first interface to beam simulations has been established that will allow visualisation of models, currently loaded optics and direct feedback from beam instrumentation. In the future, further upgrades are envisaged, reaching the full capabilities due to the software migration project, the North Area Consolidation Project, and the new CERN GUI Strategy.

\section{ACKNOWLEDGEMENTS}
The authors warmly thank G.L. D'Allesandro, D. Walter, I. Perez, M. van Dijk, M. Rosenthal, and E. Montbarbon for their important contributions to the software migration and the CERN management for their continuous support of these activities.
%
% only for "biblatex"
%
\ifboolexpr{bool{jacowbiblatex}}%
	{\printbibliography}%
	{%
	% "biblatex" is not used, go the "manual" way

} % end \ifboolexpr
%
% for use as JACoW template the inclusion of the ANNEX parts have been commented out
% to generate the complete documentation please remove the "%" of the next two commands
% 
%\newpage

%\include{annexes-A4}

\end{document}